\documentclass[twocolumn]{article}
\usepackage{graphicx}
\usepackage{amsmath}
\usepackage{amstext}
\usepackage{amssymb}

\def\thebibliography#1{\section*{\normalsize \bf References 
 }\list
 {[\arabic{enumi}]}{\settowidth\labelwidth{[#1]}\leftmargin\labelwidth 
 \advance\leftmargin\labelsep 
 \usecounter{enumi}} 
 \def\newblock{\hskip .11em plus .33em minus .07em} 
 \sloppy\clubpenalty4000\widowpenalty4000 
 \sfcode`\.=1000\relax}

\renewcommand{\Im}{{\rm Im}}

\begin{document} 
 
\twocolumn[

\hfill \fbox{\small PREPRINT: \hspace{9cm} \hfill Date: \today} \hfill 
 
\begin{center} \LARGE  
  Ferromagnetism and disorder: A dynamical mean-field study
\end{center} 
 
\begin{center} \large 
   D.\ Meyer
\end{center} 
 
\begin{center} \small \it  
  Department of Mathematics, Imperial College,
  180 Queen's Gate, London SW7 2BZ, United Kingdom
\end{center} 
\vspace{10mm} 
 
\small

---------------------------------------------------------------------------------------------------------------------------------------------------
 
\noindent{\bf Abstract} \\ 
We investigate ferromagnetism in the periodic Anderson model with
diagonal disorder.
Using 
dynamical mean-field theory in combination with the modified
perturbation theory, the disorder can be included in the calculation
consistently, which turns out to be equivalent to the CPA method.
Disorder generally reduces the Curie temperature and can for certain
configurations completely suppress ferromagnetic order. This can be ascribed to the
enhanced quasiparticle damping and the special structure of the density
of states.
{\bf PACS}\\ 
71.10.Fd, 71.28+d, 75.30.Md 
\vspace{2mm} 
 
\noindent{\bf Keywords}\\ 
Periodic Anderson Model, Ferromagnetism, Disorder, Dynamical Mean-Field Theory
 
\vspace{2mm}

---------------------------------------------------------------------------------------------------------------------------------------------------

\vspace{12mm} 
]

The investigation of models for strongly correlated electron systems has
made significant progress in recent years due to the introduction of the
dynamical mean-field theory~\cite{PJF95,GKKR96} which is based on the
non-trivial limit of infinite spacial dimensions~\cite{MV89,Met89}. In
this limit, the lattice self-energy becomes wave-vector independent and
the problem can be mapped onto a single-site
problem~\cite{Ohk92}. Many questions concerning strongly correlated
electron systems such as the Mott-Hubbard metal-insulator transition could
be answered by this approach~\cite{GK92,RZK92}. Also important insight
into the physics of band-ferromagnetism could be
gained~\cite{VBHK01,NPHW01}.

The treatment of disorder also simplifies in the
limit of infinite spatial dimensions~\cite{VV92}: The well-known CPA method~\cite{VKE68}, which has
to be seen as the best single-site approximation for solving
disorder problems~\cite{NolBd7}, becomes exact here.

As first shown by Ulmke \textit{et al.}~\cite{UJV95}, the 
dynamical mean-field theory therefore allows to investigate the interplay
of disorder and strong electron correlations by taking into account both
problems on the same level of approximation (see also ~\cite{WSC96,LCM99pre2,LF01pre}).

In this paper we want to focus on the influence of disorder on
band-ferromagnetism, in particular the ferromagnetic phase of the
periodic Anderson model in the intermediate-valence
regime~\cite{MN00c,MN00d,BBG01pre}. 
The
periodic Anderson model is defined by its Hamiltonian:
\begin{equation}
  \label{hamiltonian}
    \begin{split}
      H =&\sum_{\vec{k},\sigma}
    \epsilon(\vec{k}) s_{\vec{k}\sigma}^{\dagger}s_{\vec{k}\sigma} + 
        V \sum_{i,\sigma} (f_{i\sigma}^{\dagger}s_{i\sigma} +
    s_{i\sigma}^{\dagger}f_{i\sigma} ) +\\
     & +\sum_{i,\sigma} \epsilon_{\rm f} f_{i\sigma}^{\dagger}f_{i\sigma}
     + \frac{1}{2} U \sum_{i,\sigma}
    n^{\rm (f)}_{i\sigma}n^{\rm (f)}_{i-\sigma}
  \end{split}
\end{equation}
Here, $s_{\vec{k}\sigma}$ ($f_{i\sigma}$) and
$s_{\vec{k}\sigma}^{\dagger}$ ($f_{i\sigma}^{\dagger}$) are the
creation and annihilation operators for a conduction electron with
Bloch vector $\vec{k}$ and spin $\sigma$ (a localized electron on site
$i$ and spin $\sigma$) and
$n_{i\sigma}^{\rm (f)}=f_{i\sigma}^{\dagger}f_{i\sigma}$
($s_{\vec{k}\sigma}=\frac{1}{N}\sum_{\vec{k}} e^{i \vec{k} \vec{R}_i}
s_{i\sigma}$).
The dispersion of the 
conduction band is $\epsilon(\vec{k})$ and $\epsilon_{\rm f}$ is the
position of the localized level.
The hybridization strength $V$ is taken to be
$\vec{k}$-independent, and finally $U$ is the on-site Coulomb
interaction strength between two $f$-electrons. Throughout this paper, the
conduction band will be described by a free (Bloch) density of states,
$\rho_0(E)= \frac{1}{N}\sum_{\vec{k}} \delta (E-\epsilon(\vec{k}))$,
of semi-elliptic shape. Its width $W=1$ sets the
energy scale, and its center of gravity the energy-zero: 
$T_{ii}=\frac{1}{N}\sum_{\vec{k}}\epsilon(\vec k)\stackrel{!}{=}0$.
To obtain the single-electron Green's function for this model, we apply
dynamical mean-field theory (DMFT) in combination with the modified
perturbation theory (MPT)~\cite{MWPN99} to solve the associated impurity
problem. The DMFT approach becomes exact for 
$Z\rightarrow\infty$ ($Z$ being the coordination number) and represents
a well-defined local approximation for finite dimensions (finite $Z$).
The DMFT-MPT approach recovers the high-energy features of the Green's
function up to order $(\frac{1}{E^4})$. Furthermore, its low-energy
behaviour is at least qualitatively correct. The investigation of
ferromagnetism in the Hubbard model~\cite{PHWN98,Ulm98} as well as the
Mott-Hubbard transition~\cite{GKKR96,BCV01} have shown that this method
is able to predict qualitatively correct phenomena of strongly
correlated systems.
Numerical results are obtained using a standard DMFT algorithm which
basically consists of a self-consistency loop as follows:
Starting with an initial guess for the self-energy for the lattice, the
conduction electron bath of an associated impurity problem is defined
using the \textit{self-consistency equation}~\cite{GKKR96}. Then this
impurity problem is solved by some means, and its self-energy
extracted. The latter is then taken to be the lattice self-energy and a
new impurity model is defined via the self-consistency equation. This loop is iterated until
self-consistency is achieved. More details on practical calculations
can be found in ~\cite{GKKR96,MeyerDiss}. 

Next we need to specify how the
impurity model is solved. For this we employ the modified perturbation
theory. This method is based on the following ansatz for the
self-energy~\cite{MFBP82,KK96}:
\begin{equation}
  \label{eq:ansatz}
  \Sigma_{\sigma}(E)=U \langle n_{-\sigma}^{(f)}\rangle
  +\frac{\alpha_{\sigma} \Sigma_{\sigma}^{\rm (SOC)}(E)}
  {1-\beta_{\sigma} \Sigma_{\sigma}^{\rm (SOC)}(E)}
\end{equation}
$\alpha_\sigma$ and $\beta_{\sigma}$ are introduced as parameters to be
determined
later. $\Sigma_{\sigma}^{\rm (SOC)}(E)$ is the
second-order contribution to perturbation theory around the Hartree-Fock 
solution~\cite{Yam75}. 
Equation~(\ref{eq:ansatz}) can be understood as the simplest possible
ansatz which can, on the one hand, reproduce the perturbational result
in the limit $U\rightarrow 0$, and, on the other hand, recovers the
atomic limit for
appropriately chosen $\alpha_{\sigma}$ and $\beta_{\sigma}$~\cite{MFBP82}.

Using the perturbation theory around the Hartree-Fock solution
introduces an ambiguity into the calculation. Within the self-consistent
Hartree-Fock calculation, one can either choose the chemical potential
to be equivalent to the chemical potential of the full MPT calculation,
or take it as parameter $\tilde{\mu}$ to be fitted to another physically
motivated constraint.
In reference~\cite{KK96} the
Luttinger theorem~\cite{LW60}, or equivalently the Friedel sum
rule~\cite{Fri56,Lan66}, was used to determine $\tilde{\mu}$.
As discussed in  Ref.~\cite{MN00c},
we use the physically motivated condition of
identical impurity occupation numbers for the Hartree-Fock and 
the full calculation ($n_{\sigma}^{(f,{\rm HF})}=n_{\sigma}^{(f)}$) to
determine $\tilde{\mu}$, which also allows for a consistent extension of the
method to finite temperatures~\cite{MN00b,MN00c}. Except for symmetric
parameters this will lead to an approximate
fulfillment of the Luttinger theorem only~\cite{MWPN99}. 
Bearing in mind that for disordered systems, the Luttinger theorem does
not apply, this should not be a decisive disadvantage for this study.
A more detailed analysis of the
different possibilities to determine $\tilde{\mu}$ is found in
reference~\cite{PWN97} where the DMFT-MPT was applied to the single-band
Hubbard model.
Finally, the parameters $\alpha_{\sigma}$ and $\beta_{\sigma}$
have to be determined. Instead of using the ``atomic'' limit of $V=0$ as 
was done for example in references~\cite{MFBP82,MLFT86,KK96}, we
make use of the moments of the spectral density. This procedure is
described in detail in references~\cite{PWN97,MWPN99}. The result
not only
fulfills the $V=0$ limit, but also recovers the high-energy behavior of
the Green's function up to the order $(\frac{1}{E^4})$.

\begin{figure}[t]
  \begin{center}
    \includegraphics[width=6cm]{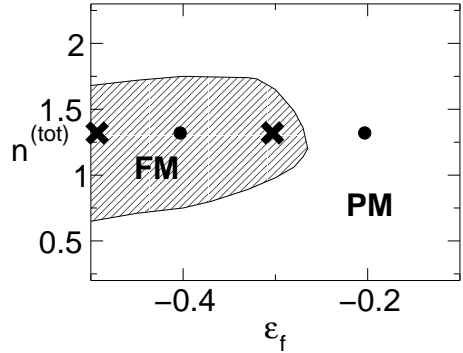}
    \caption{Zero-temperature phase diagram for the periodic Anderson
      model without disorder from ~\cite{MN00c}, $U=4$ and
      $V=0.2$. In the shaded region, the system is ferromagnetic,
      elsewhere paramagnetic.}
    \label{fig:phdia}
  \end{center}
\end{figure}

The results concerning ferromagnetism in the periodic Anderson model
have been discussed elsewhere~\cite{MN00c,MN00d,MeyerDiss}. For the
following it is important to note that in the intermediate-valence
regime, more precisely for $-W/2 \lesssim \epsilon_{\rm f} \lesssim  -W/4$,
a ferromagnetic solution with finite $T_{\rm c}$ exists for a range
of electron densities. The $T=0$ phase diagram is plotted in
Fig.~\ref{fig:phdia}. In this area of the phase diagram, the ferromagnetic solution
shows typical features of a band-ferromagnet~\cite{VBHK01,NPHW01}. The origin of the
ferromagnetic order lies in the competition of kinetic and potential
energy. Ferromagnetic order is stabilized by high values of the
density of states (DOS) close to the lower band
edge. The electron density needs to be chosen that the
chemical potential lies in this region of large DOS. So although Stoner's
theory~\cite{Sto36} does not capture the right physics, his criterion for
the occurrence of band-ferromagnetism ($U \rho(\mu) \gg 1$) turns out to
remain valid~\cite{VBHK01,NPHW01}.

Let us now turn to the problem of disorder. 
A general extension of model~(\ref{hamiltonian}) to include
diagonal (on-site) disorder is
\begin{align}
  \label{eq:disorder_in_ham}
  H \rightarrow &H+H^{\rm (dis)}\\
  \begin{split}
    &H^{\rm (dis)} =   \sum_{i,\sigma} \Delta V_i (f_{i\sigma}^{\dagger}s_{i\sigma} +
    s_{i\sigma}^{\dagger}f_{i\sigma} )+\\
    &\quad+\sum_{i,\sigma} \Delta\epsilon_{\rm f,i} f_{i\sigma}^{\dagger}f_{i\sigma}
    + \frac{1}{2}  \sum_{i,\sigma} \Delta U_i
    n^{\rm (f)}_{i\sigma}n^{\rm (f)}_{i-\sigma}
  \end{split}
\end{align}
In this model, the $f$-electron energy ($\epsilon_{\rm f}$), the on-site
hybridization ($V$) and the interaction strength ($U$) can deviate from
the value denoted before by $\Delta\epsilon_{f,i}$, $\Delta V_i$ and
$\Delta U_i$, respectively. The distribution for each of these quantities
can be defined by a probability distribution function
$P(\Delta\epsilon_{f,i})$, $P(\Delta V_i)$ and $P(\Delta U_i)$.

A standard method to solve electron systems with disorder is the
well-known \textit{coherent potential approximation}
(CPA)~\cite{VKE68}. This method is considered the best single-site
approximation and at least for one-particle properties, has proven to be
remarkably successful~\cite{VV92}. 
It is known to become exact in a number of
limiting cases, namely for small impurity concentration, small potential
strengths, vanishing inter-site hopping, and as discussed by~\cite{SS72,VV92}
also for $Z^{-1}$.
This last limiting case suggests that CPA results could be obtained by
an alternative (DMFT-like) algorithm employing a mapping onto an
impurity model. This was shown to be the case~\cite{UJV95}, and allows for a systematic extension to include
many-body interactions, which would be not possible within the standard
CPA procedure~\cite{UJV95,Byzcukprep}.
One has to bear in mind, however, that single-site approximations such as the
one described here do have some limitations. A
major limitation in regards to disorder is the inability to describe
inhomogeneous, phase-separated systems which are
discussed in the context of manganites~\cite{DHM01}.

To calculate the single-electron Green function for a (diagonally) disordered system
using DMFT, the algorithm described above for pure systems needs to be
modified in the following way:
Instead of one, several impurity models need to be solved, one for each
possible on-site configuration of the lattice model. The configurational
averaging is then performed on the results for these impurity models, and the averaged self-energy
extracted and taken as lattice self-energy. This self-energy is fed into the
self-consistency equation to determine a new impurity bath
function. This cycle is then iterated until self-consistency is
obtained~\cite{UJV95,Byzcukprep}.

In the following, we want to present and discuss numerical results for
the periodic Anderson model with diagonal disorder. For simplicity, we
have set all $\Delta U_i = 0$ and $\Delta V_i=0$, including only
disorder with respect to the $f$-level energy $\epsilon_{\rm f}$. We
further restrict ourselves to binary alloys, i.e. $P(\Delta
\epsilon_{\rm f,i}) = p \delta(\Delta \epsilon_{\rm f}) + (1-p)
\delta(0)$. The in general rather complex definition of the disorder is
reduced to two parameters: the difference of $f$-level positions of the
two components $\Delta\epsilon_{\rm f}$, and the concentration $p$. 

\begin{figure}[t]
  \begin{center}
    \includegraphics[width=0.4\textwidth]{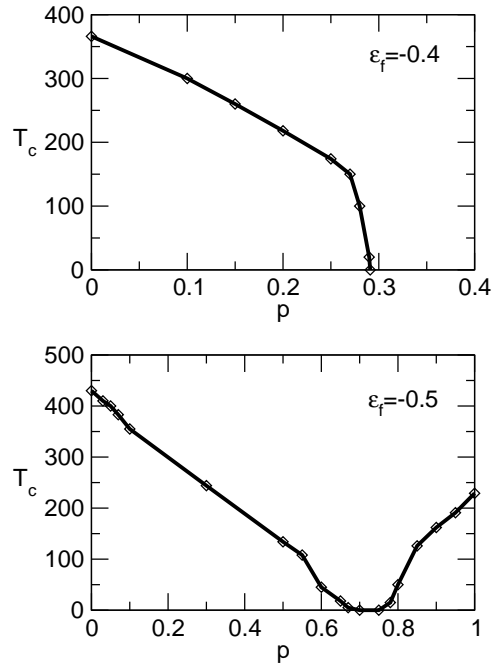}
    \caption{Curie temperatures for the PAM with disorder for $U=4$,
      $n^{\rm (tot)}=1.3$, $V=0.2$. The $\epsilon_{\rm f}$ and the disorder
      parameters are explained in the text.}
    \label{fig:tc}
  \end{center}
\end{figure}
\begin{figure}[ht]
  \begin{center}
    \includegraphics[width=0.45\textwidth]{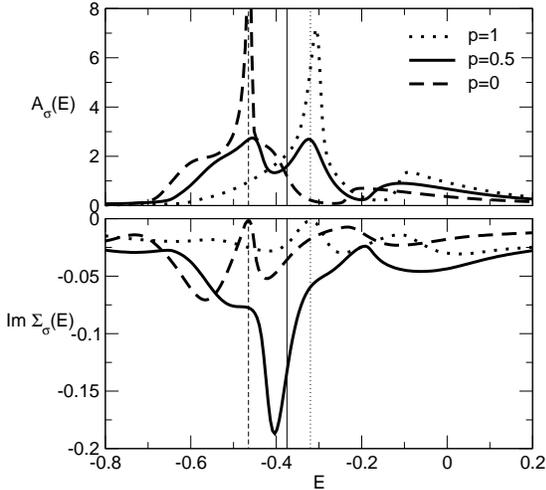}
    \caption{Density of states $A_\sigma(E)=-\frac{1}{\pi} \Im G^{\rm
        (f)}_\sigma(E)$ and $\Im \Sigma_\sigma(E)$ for small interaction
      strength $U=0.2$, $\epsilon_{\rm f}=-0.5$, $\Delta \epsilon_{\rm
        f}=0.2$, $V=0.2$ and $n^{\rm (tot)}=1.3$ for the two pure limits
      $p=0$ and $p=1$ as well as for $p=0.5$}
    \label{fig:sigdos}
  \end{center}
\end{figure}

To examine the influence of disorder on ferromagnetic order, we investigate
two different scenarios using this simple model of disorder:
Selecting $U$, $V$ and the electron density $n^{\rm (tot)}$ so that a
ferromagnetic solution is possible for a range of values of
$\epsilon_{\rm f}$, we can choose $\epsilon_{\rm f}$ and
$\Delta \epsilon_{\rm f}$ so that one of the two alloy components would
in a pure system be ferromagnetic, and the other not. 
We have taken $\epsilon_{\rm f}=-0.4$ and $\Delta
\epsilon_{\rm f} = 0.2$. These parameters for the two alloy
components are indicated in the
phase diagram (Fig.~\ref{fig:phdia}) as solid circles. 
The other scenario
corresponds to alloying two different ferromagnetic materials. 
It is realized by using $\epsilon_{\rm f}=-0.5$ and $\Delta\epsilon_{\rm f}=0.2$. The resulting
components are shown by the crosses in Fig.~\ref{fig:phdia}.

Let us start by looking at the first scenario: For $p=0$ the system is
ferro-, for $p=1$ paramagnetic. In Fig.~\ref{fig:tc}, we have plotted the Curie
temperature as function of $p$. As one would expect, $T_c$ decreases  with
increasing $p$, in the beginning linearly, and around $p\approx 0.28$,
rather suddenly. With $p\gtrsim 0.29$, the system is paramagnetic already for $T=0$.

The second scenario reveals more unexpected behaviour. For $p=0$ as well
as for $p=1$ (corresponding to the two crosses in Fig.~\ref{fig:phdia}) the
system is ferromagnetic and has finite $T_c$. 
In between these two pure limits, we find a strong reduction of the Curie
temperature. Around $p\approx 0.7$, $T_c$ even becomes
zero and the system
is then paramagnetic. The reduction of $T_c$ can be ascribed to the
additional quasiparticle damping induced by the disorder. This effect of
quasiparticle damping was already
noticed in previous studies investigating the role of quasiparticle
damping in the Hubbard model~\cite{HN96} and in the PAM~\cite{MN00d}. The
cited works did not involve disorder, but tested the influence of
quasiparticle damping by comparing different approximation schemes some
of which neglected quasiparticle damping completely. A similar reduction
of $T_c$ due to disorder was also found for a ferromagnetic
Kondo-lattice model with classical spins~\cite{LF01pre}.

To display the enhanced quasiparticle damping due to disorder, we have performed calculations for a
weakly interacting ($U=0.2$), paramagnetic system. In Fig.~\ref{fig:sigdos}, the
resulting f-density of states is plotted together with the
imaginary part of the self-energy. These two quantities were calculated
for $p=0$, $p=0.5$ and $p=1$. Except for the smaller value of $U$, all
other parameters were taken as in the lower panel of
Fig.~\ref{fig:tc}. For both pure situations, the self-energy vanishes quadratically
at the Fermi energy which is indicated by the respective thin vertical line
(note: the chemical potential is shifted by varying $p$ since we keep
the total electron density $n^{\rm (tot)}$ constant). In the alloyed
compound, the imaginary part of the self-energy remains finite at the
Fermi energy. And over a large energy range 
it is strongly enhanced compared to the pure limits. 
This strong disorder-induced quasiparticle damping is independent of
the interaction strength, and can have a suppressing effect
on ferromagnetism~\cite{MN00d}.

Another effect is complementing the quasiparticle damping as mechanism to
completely suppress ferromagnetism as seen in Fig.~\ref{fig:tc}. As
already mentioned before, the occurrence of band-ferromagnetism is linked
to a Stoner-like criterion requiring a large density of states at or
close to the Fermi energy~\cite{NPHW01,VBHK01}.
For both pure cases ($p=0$ and $p=1$), the Fermi energy lies within
the charge excitation peak in the density of states as can be seen from
Fig.~\ref{fig:sigdos} for the small $U$ case. In the disordered case,
however, the structure of the DOS is dominated not by one, but two
charge excitation peaks. The positions of these are given by the
respective charge excitations in the pure limits. For intermediate
values of $p$ ($p \approx 0.5$), the Fermi energy lies in between these two peaks, and
the value of the density of states at this energy is relatively
low. This should therefore reduce the tendency towards ferromagnetism.
The special structure of
the DOS for intermediate $p$ leads therefore to further suppression of
ferromagnetism. Whether (and for which $p$) $T_c$ really vanishes, or not, now critically
depends on the position of the Fermi energy, and therefore on the
electron density.

To summarize, we have performed dynamical mean-field theory calculations
for a periodic Anderson model (PAM) with disorder. The inclusion of disorder
into these calculations follows a relatively simple and straightforward
recipe~\cite{UJV95,Byzcukprep}. Our numerical analysis shows how
even small amounts of disorder can reduce the Curie temperature of a
ferromagnetically ordered PAM in the 
intermediate-valence regime significantly. Stronger disorder can lead to a
complete suppression of the ferromagnetic phase, even for a binary alloy
of two ferromagnetically ordered components.

\noindent{\bf Acknowledgements}\\
The author wants to thank K. Byczuk for introducing me to this
topic, and D. Edwards, A. Hewson and W. Nolting for pleasant and helpful
discussions.

\end{document}